\documentclass[10pt,letterpaper]{article}
\usepackage[top=0.85in,left=2.75in,footskip=0.75in]{geometry}

\usepackage{amsmath,amssymb}

\usepackage{changepage}

\usepackage{textcomp,marvosym}

\usepackage{cite}

\usepackage{nameref,hyperref}

\usepackage[right]{lineno}

\usepackage[nopatch=eqnum]{microtype}
\DisableLigatures[f]{encoding = *, family = * }

\usepackage[table]{xcolor}

\usepackage{array}

\newcolumntype{+}{!{\vrule width 2pt}}

\newlength\savedwidth

\raggedright
\usepackage[aboveskip=1pt,labelfont=bf,labelsep=period,justification=raggedright,singlelinecheck=off]{caption}

\makeatletter
\renewcommand{\@biblabel}[1]{\quad#1.}
\makeatother

\usepackage{lastpage,fancyhdr,graphicx}
\usepackage{epstopdf}
\fancyheadoffset[L]{2.25in}
\fancyfootoffset[L]{2.25in}
\begin{document}
\vspace*{0.2in}

\begin{flushleft}
{\Large
\textbf\newline{Affective publics in Arabic YouTube}
}
\newline
\\
Lynnette Hui Xian Ng\textsuperscript{2*},
Craig Douglas Albert\textsuperscript{1*},
Abdullah Melhem\textsuperscript{3},
Ahmed Aleroud\textsuperscript{3},
Lance Y.\ Hunter\textsuperscript{1}
\\
\bigskip
\textbf{1} Department of Social Sciences,
Augusta University, Augusta, Georgia, USA
\\
\textbf{2} Societal Computing, Carnegie Mellon University,
Pittsburgh, Pennsylvania, USA
\\
\textbf{3} School of Computer and Cyber Sciences,
Augusta University, Augusta, Georgia, USA
\\
\bigskip
* Corresponding author: lynnetteng@cmu.edu; calbert@augusta.edu

\end{flushleft}

\section*{Abstract}
What is the emotional register of Arabic YouTube's affective publics? To investigate this, we analyzed 67,725 YouTube comments collected around socio-political topics
associated with Yemen, Saudi Arabia, Iraq, Jordan, and Syria using a unified
sentiment-and-emotion pipeline. Our results profile a single regional affective public rather than five separate national ones. Sentiment is overwhelmingly negative across all five country-oriented corpora, and the country-level emotion profiles are structurally similar. This shared register still accommodates some regional variations: discourse is organized around country-level political actors and cross-border historical trauma figures, and grief singularizes Iraq from the other countries. The differences in emotional register also tracks lived political causes rather than fixed categories, which we observe from patterns of the valence of US-related content, that tracks the presence or absence of direct US military engagement.
Our work shows that the emotional register of Arabic YouTube's affective publics is a shared one that is historically layered and geographically conditioned, which has implications for public diplomacy in the MENA region.

%\linenumbers

% ======================================================================
\section{Introduction}
% ======================================================================
Arabic-language social media is a primary site of political expression in
the Middle East and North Africa (MENA) \cite{lynch2011after}, but existing
computational work has not answered this question at scale. Past work remains
confined largely to Twitter/X, typically reports only binary sentiment
classifications (positive/negative) or single-country designs
\cite{ref4,ref5,ref8,ref9}, and rarely asks which specific emotions organize
the negativity it detects, whether that register holds together across
national borders, or how it shifts when discourse turns to a shared external
actor like the United States.

This study addresses that gap by moving from the text-based platform
Twitter/X to YouTube, the dominant video platform in the region (eight of
the world's top thirteen countries by YouTube usage are MENA states
\cite{ref15}), and measuring the emotional register directly rather than
inferring it from sentiment alone. We use 67,725 Arabic YouTube comments
from Yemen, Saudi Arabia, Iraq, Jordan and Syria, collected during the
summer of 2024, and apply a unified computational pipeline combining
CAMeL-Lab BERT for Arabic sentiment classification and EmoRoBERTa for
28-category emotion detection, with machine translation as an intermediate
step for emotion analysis. We also examine how the presence of US-related
content affects this register, and whether that relationship varies across
countries.

Our study asks three research questions organized around this register:
\textbf{(RQ1:)} How do sentiments expressed in Arabic YouTube comments on
socio-political topics differ across the five countries? \textbf{(RQ2:)}
What is the emotional register underneath that sentiment? This question studies the emotions that
are most commonly expressed, and how the register differ across the five
national contexts. \textbf{(RQ3:)} How does this register shift in the
presence of content related to the United States (US), and does that shift vary
across countries?

The study contributes to the literature in three aspects. First, it
provides a cross-national mapping of the emotional register in
Arabic YouTube comments across five MENA countries using a unified
computational pipeline. Second, it shows that the emotional register itself, and not aggregate sentiment, carries the more diagnostic signal of the publics; that is, communal solidarity emotions dominate throughout, an interpretation invisible from
sentiment scores alone. Third, it documents that this emotional register shifts
systematically with US-related content in a way that is geopolitically
differentiated by country, with discussion of the implications for US public
diplomacy strategy in the region.

% ======================================================================
\section{Background and Related Work}
% ======================================================================

The digital publics in MENA countries operate under unique conditions that
differ markedly from Western contexts in which many computational political
communication studies have been conducted. In particular, the MENA digital
publics operate under conditions of authoritarian constraint, active
information warfare, state propaganda, chronic political instability, and
high platform penetration relative to formal democratic participation
\cite{ref1,ref13,ref20,ref28}. YouTube is a primary political arena for
hundreds of millions of Arabic speakers, with eight of the world's top
thirteen countries by YouTube usage in the MENA region \cite{ref15}.

Research on social media and political affect in MENA has established
several baseline findings: negativity predominates across platforms and
countries \cite{ref4,ref5}; the US military presence and the
 dynamics of the Israel-Palestine conflict are consistent drivers of negative sentiment
\cite{ref6,ref7}; and anti-Americanism scales with the intensity of US
military and economic intervention rather than remaining constant across
contexts \cite{glas2021antiamericanism}. Cross-national comparisons, such as
Jamal et al.'s \cite{jamal2015antiamericanism} analysis of Arabic Twitter
discourse spanning Egypt and Syria, typically only report aggregate sentiment
(e.g., a roughly 3:1 ratio of negative to positive tweets), and do not
examine whether negative sentiment is organized by communal, grief-driven, or
solidarity-oriented emotions.

Computational work on Arabic sentiment has grown substantially over the past
decade \cite{ref3}, moving from lexicon-based and classical machine-learning
approaches applied mostly to product reviews and news text toward transformer
models pre-trained on large Arabic corpora, with applications spanning social
and political commentary \cite{ref2,ref4,ref5} and Twitter-based political
discourse during regional crises \cite{ref6,ref7}. Most of this work, though,
remains confined to binary sentiment classification within a single country
\cite{ref8,ref9}. Emotion analysis is considerably less developed: existing emotion detection models still grapple with the nuances of differentiation \cite{ng2021bot}; and existing
Arabic emotion corpora cover narrow ground, including Abdullah et al.'s
\cite{ref67} extraction of emotions from Arabic tweets and Al-Mahdawi and
Teahan's \cite{ref52} six-category annotation of an Iraqi Arabic Facebook
corpus, and neither approaches the scale of English-language benchmarks such
as GoEmotions, which spans 58,000 Reddit comments labeled across 28
categories \cite{demszky2020goemotions}. The closest precedent for
cross-national comparison is Al-Laith and Shahbaz \cite{ref53}, who built a
temporal Arabic news sentiment dataset spanning multiple Arab states using a
unified computational approach and found Iraq to be the most negative country
in their corpus (88\% negative news coverage).

Much of this computational apparatus was built and validated on English
text, and porting it to Arabic is not a trivial task: Arabic is
morphologically complex, dialects vary widely across regions, and
comparatively little labeled social media data exists to train on, so tools
developed for English do not transfer cleanly \cite{ref3}. The gap is
sharper for emotion than for sentiment, since no Arabic-language resource
matches the scope of English benchmarks like GoEmotions, forcing emotion
analysis through machine translation as an intermediate step and risking the
loss of culturally specific emotional vocabulary with no clean English
equivalent---a limitation we return to later in the paper. Against this
backdrop, the present study extends prior work along three dimensions: it
incorporates fine-grained, 28-category emotion detection alongside
binary-valence sentiment rather than sentiment alone; it moves the analysis
from Twitter/X, where most prior Arabic work is concentrated
\cite{ref4,ref5,ref6,ref7}, to YouTube, the region's dominant video platform;
and it compares five national corpora within a single unified computational
pipeline rather than the single-country or aggregate-only designs that have
characterized this literature to date.

% ======================================================================
\section{Materials and Methods}
% ======================================================================

\subsection{Data collection}

We collected Arabic YouTube comments during the summer of 2024 using the
\texttt{youtube\_comment\_downloader} and \texttt{youtubesearchpython} Python
libraries. Searches were conducted using country-specific Arabic keywords
relating to socio-political issues relevant to each national context, such as the war in Gaza, domestic political conflicts, armed conflict and US involvement (see \autoref{tab:keywords} for themes). Accordingly, country labels refer to the national context around which each corpus was collected rather than the verified nationality or geographic location of individual commenters. Cross-country differences are therefore interpreted as differences among country-oriented discourse corpora, not national populations. The full keyword list with Arabic terms and English translations is provided in Supporting Information \nameref{supp:keywords}). The resulting dataset contained 67,725 comments: Yemen ($n = 19{,}020$),
Saudi Arabia ($n = 19{,}001$), Iraq ($n = 10{,}321$), Jordan ($n = 10{,}273$),
and Syria ($n = 9{,}110$). Each record includes video title, duration, view
count, description, comment text, sentiment scores, emotion scores, named entity
recognition output, and a binary indicator for US mentions which was added on as a post-processing step by text-matching identification
(\texttt{Contains\_USA}). 

\begin{table}[!ht]
\caption{
{\bf Thematic domains of search keywords by country.}
Full keyword list with Arabic terms is provided in Supporting Information
\nameref{supp:keywords}.}
\label{tab:keywords}
\begin{tabular}{lp{3.8in}}
\hline
\textbf{Country} & \textbf{Thematic Domains} \\
\hline
Yemen & Houthi conflict, Saudi coalition, US support, Gaza solidarity, Yemeni governance \\
Saudi Arabia & Gaza war, Vision 2030, US--Saudi relations, regional security \\
Jordan & Gaza proximity, Palestinian cause, US policy, Jordanian political protests \\
Iraq & US forces in Iraq, ISIS remnants, Iranian influence, Iraqi governance, Gaza \\
Syria & Syrian civil war, US forces in northeast Syria, reconstruction, displacement \\
\hline
\end{tabular}
\end{table}

\paragraph{Ethics statement.}
This study analyzes publicly available YouTube comments. No personally
identifying information was collected or retained; all text was analyzed at the aggregate level. Data collection was conducted in accordance with
YouTube's Terms of Service for research purposes.

\subsection{Country contexts}

The five national contexts differ enough in history, platform penetration,
and relationship to the United States that brief background is necessary for
interpreting the results that follow. In Yemen, which has almost no existing
Arabic sentiment analysis literature, social media activity has been shaped
by the Houthi-Saudi conflict since 2015 \cite{ref20,ref21}; US involvement
there is indirect, channeled through support for the Saudi-led coalition
rather than a direct military presence \cite{ref19,ref25}. Saudi Arabia, by
contrast, has among the highest platform penetration in the sample (35
million social media users and 28 million YouTube users as of 2024
\cite{ref27}) within a media environment that is highly state-aligned, with
strong censorship and surveillance \cite{jones2022digital}; its relationship
with the United States is a strategic alliance in which the US is a
significant geopolitical actor in Saudi economy and foreign policy but, as in
Yemen, does not maintain a direct military presence \cite{ref31}. Jordan,
with 6.38 million active social media users \cite{ref34}, faces intense
affective pressure from its geographic proximity to the Gaza conflict and a
historically complex relationship with both the US and Israel
\cite{ref36,ref39}; existing Jordanian Arabic sentiment research finds
predominantly negative sentiment on political topics \cite{ref41}.

Iraq and Syria present a different configuration. Iraq has 31.95 million
social media users, of whom 22.8 million use YouTube \cite{ref43}, and holds
the most developed Arabic emotion-analysis literature of the five countries
\cite{ref51,ref52,ref53}; Al-Laith and Shahbaz \cite{ref53} rank it the most
negative country in their multi-country corpus (88\% negative news). The US
maintained a direct military presence in Iraq from 2003 to 2011 and returned
in 2014 to support Iraqi forces against ISIS, though by 2026 those forces are
beginning to withdraw. Syria presents the most constrained data environment
of the five: over 60\% of the population lacks internet access \cite{ref54},
meaning the Syrian corpus likely overrepresents urban, diaspora, and
digitally connected populations. US forces operate in northeastern Syria, and
the relationship is experienced as a direct military presence, though at
lower intensity than in Iraq.

\subsection{Computational pipeline}
After collection of the YouTube video comments, we run through a sentiment-and-emotion classification pipeline, which we detail in the following section.

\paragraph{Sentiment classification.}
Sentiment was classified using \textbf{CAMeL-Lab BERT}\footnote{\url{https://huggingface.co/CAMeL-Lab/bert-base-arabic-camelbert-da-sentiment}}, which is a transformer
model fine-tuned on dialectal Arabic social media text \cite{ref72,ref73}. To
handle Arabic's complex morphological context, we employed a trigram-based
scoring approach. For a comment $C$ with $N$ tokens $w_1, w_2, \ldots, w_N$,
overlapping trigrams are extracted as:

\begin{equation}
G_i = (w_i,\; w_{i+1},\; w_{i+2}), \quad i = 1, 2, \ldots, N - 2
\label{eq:trigram}
\end{equation}

Each trigram receives a sentiment score $S(G_i)$ from CAMeL-Lab BERT. The
comment-level positivity score is the mean trigram score:

\begin{equation}
\bar{S}(C) = \frac{1}{N-2} \sum_{i=1}^{N-2} S(G_i)
\label{eq:sentiment}
\end{equation}

Neutrality is computed as a model confidence measure per trigram, averaged
across the comment. All Positivity, Negativity, and Neutrality scores are on a
0--100 scale. Averaging over overlapping trigrams reduces sensitivity to any
single token's contribution, which is important for dialectal Arabic where
polarity can shift at the sub-sentence level.

\paragraph{Emotion classification.}
Emotion was classified using \textbf{EmoRoBERTa}\footnote{\url{https://huggingface.co/arpanghoshal/EmoRoBERTa}}, which we then fine-tuned on the GoEmotions dataset for a finer-grained 28-category emotion classification \cite{ref67,ref68,ref69,ref70}. Because
EmoRoBERTa was trained on English text, we translated the Arabic trigrams that we extracted in the previous step into English using the Google Translate machine translation model from the deep-translator Python package\footnote{\url{https://deep-translator.readthedocs.io/en/latest/}}. The translated English texts were checked by a native Arabic speaker, who is fluent in English at the graduate level. The resulting English trigrams were then passed to EmoRoBERTa for emotion classification, yielding a probability distribution over 28 emotion categories for each trigram. Finally, comment-level emotion scores are the sum of
per-trigram probabilities for each emotion category:

\begin{equation}
e_{i,k} = \sum_{j=1}^{n_i - 2} p_{ijk}, \quad k = 1, \ldots, 28
\label{eq:emotion_sum}
\end{equation}

where $p_{ijk}$ is EmoRoBERTa's probability for emotion $k$ on the $j$-th
trigram of comment $i$, and $n_i$ is the token count of comment $i$.

Because $e_{i,k}$ is a sum over $n_i - 2$ trigrams, comment-level scores scale
mechanically with comment length. To remove this dependence before comparing
countries, we length-normalize by the number of trigrams contributing to the
sum:

\begin{equation}
e_{i,k}^{\text{norm}} = \frac{1}{n_i - 2} \sum_{j=1}^{n_i - 2} p_{ijk}, \quad k = 1, \ldots, 28
\label{eq:emotion_norm}
\end{equation}

Comments with $n_i < 3$ cannot form a trigram (3{,}411 of 67{,}725 comments,
5.0\%) and are excluded from the length-normalized analysis. Country-level
emotion profiles are then computed as the mean of the normalized comment-level
vectors across all comments in that country:

\begin{equation}
\mathbf{E}_c = \frac{1}{N_c} \sum_{i \in c} \mathbf{e}_i^{\text{norm}}
\label{eq:country_profile}
\end{equation}

All emotion results reported below use $\mathbf{E}_c$ from
\autoref{eq:emotion_norm}--\ref{eq:country_profile}. The normalization is done because comment length differs significantly across countries ($F(4, 67{,}720)
= 9.66$, $p < 0.001$), and we want to ensure that the emotion profiles reflect the distribution of emotions rather than the distribution of comment lengths. Full distributions in \hyperref[supp:complength]{S1 Table}). 
We report length-normalized values throughout the remainder of the paper.
The dominant emotion per comment is $\hat{E}_i = \arg\max_k\; e_{i,k}^{\text{norm}}$. 

%We acknowledge that machine translation may systematically suppress emotionally
%specific Arabic vocabulary related to religious sentiment, collective mourning,
%and political solidarity; emotion results should be interpreted as relative
%cross-country comparisons rather than absolute prevalence estimates.

\paragraph{Named entity recognition and USA flag.}
Part-of-speech tagging and named entity recognition used
\texttt{hatmimoha/arabic-ner}\footnote{\url{https://huggingface.co/hatmimoha/arabic-ner}}. A binary indicator variable \texttt{Contains\_USA}
was assigned 1 for any comment in which either the comment text or its
associated video title contained a named reference to the United States (USA), in either Arabic or transliterated form. USA-mention counts by country are reported in
\autoref{tab:usa}.

\subsection{Statistical analysis}

\textbf{Sentiment differences across countries.} One-way ANOVA was used to test
whether Positivity and Negativity scores differed significantly across the five
countries. Significant ANOVA results were followed by all pairwise Welch
$t$-tests with Bonferroni correction for 10 pairwise comparisons; effect sizes
were computed as Cohen's $d$.

\textbf{Comparisons of USA mentions.} Within each country, all comment-level
sentiment score distributions violated normality (Shapiro-Wilk $p < 0.001$).
Mann-Whitney $U$ tests were therefore used to compare Positivity and Negativity
between USA-mentioned and non-USA comments within each country.

\textbf{Emotion profile similarity.} Country-level emotion profile similarity
was quantified via cosine similarity between mean 28-dimensional emotion vectors
$\mathbf{E}_c$:

\begin{equation}
\mathrm{sim}(c_1, c_2)
= \frac{\mathbf{E}_{c_1} \cdot \mathbf{E}_{c_2}}
       {\|\mathbf{E}_{c_1}\| \cdot \|\mathbf{E}_{c_2}\|}
\label{eq:cosine}
\end{equation}

%All intermediate data files are archived in the project data directory.
%Analysis code will be made publicly available upon acceptance.

% ======================================================================
\section{Results}
% ======================================================================

\subsection{Sentiment is predominantly negative and significantly
different across countries}
\label{sec:rq1}

\begin{figure}[!ht]
\centering
\includegraphics[width=0.5\linewidth]{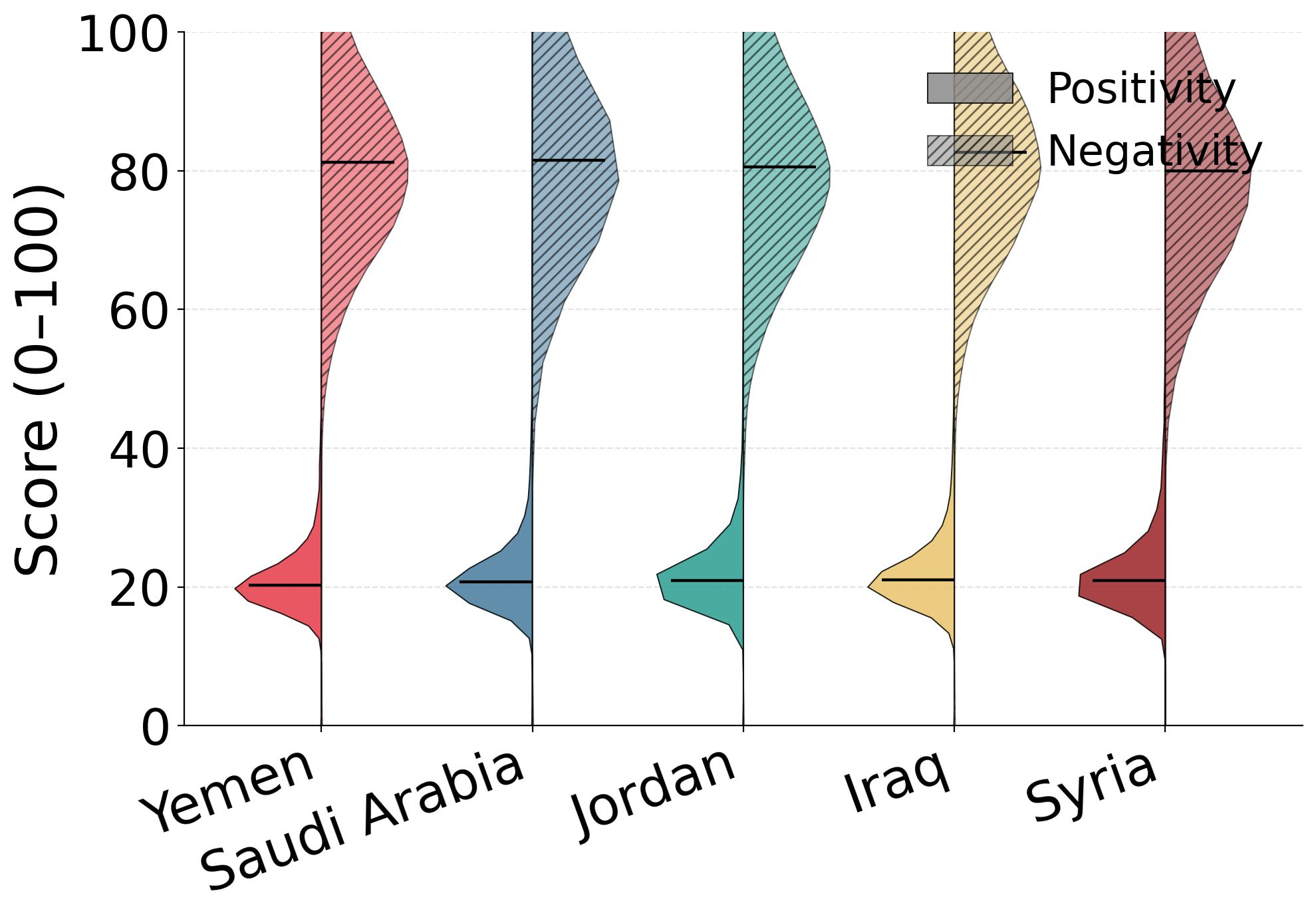}
\caption{{\bf Distribution of Positivity and Negativity scores across five
countries (0--100 scale).}
Violin width represents probability density. Horizontal
tick indicates the median. All countries exhibit predominantly negative
sentiment, concentrated at the high end of the scale, alongside a much smaller
mass of low-end Positivity scores. Iraq has the widest negativity distribution
and longest upper tail; Jordan and Syria show nearly identical distributional
shapes.}
\label{fig:violins}
\end{figure}

For RQ1, we compared the sentiment across the five country-oriented corpora (\autoref{fig:violins}, full results in \autoref{supp:sentiment}).
Across all five countries, sentiment in Arabic YouTube comments on
socio-political topics is overwhelmingly negative. Mean negativity ranges from
82.0\% (Syria) to 84.7\% (Iraq); mean positivity from 21.5\% (Yemen) to 22.4\%
(Iraq). One-way ANOVA yields significant differences across the five countries for Positivity ($F(4, 67{,}720) = 40.52$, $p < 0.001$), and also for Negativity ( $F(4, 67{,}720) = 29.36, p < 0.001$).
The raw
magnitude of negativity differences (2.7\% between Syria
(82.0\%) and Iraq (84.7\%)) is modest. This is confirmed by effect sizes: all
pairwise Cohen's $d$ values fall in the range $|d| = 0.047$--$0.132$
(\autoref{tab:pairwise}), indicating small-to-negligible practical differences.
The practical
interpretation is that all five countries share a baseline of pronounced
negativity, with Iraq measurably but modestly above the others.

\begin{table}[!ht]
\caption{
{\bf Bonferroni-corrected pairwise Welch $t$-test results for mean Negativity
(significant comparisons only; $p_{\text{Bonferroni}} < 0.05$).}
$\Delta$ Mean (pp): positive values indicate the first-named country is more
negative. Non-significant pairs: Yemen--Saudi Arabia ($p = 1.0$);
Jordan--Syria ($p = 1.0$).}
\label{tab:pairwise}
\begin{tabular}{llrr}
\hline
\textbf{Comparison} & \textbf{$\Delta$ Mean (pp)} &
\textbf{Cohen's $d$} & \textbf{$p$ (Bonferroni)} \\
\hline
Yemen vs.\ Jordan        & $+$0.86 & $+$0.047 & 0.003    \\
Yemen vs.\ Syria         & $+$1.13 & $+$0.060 & $<$0.001 \\
Yemen vs.\ Iraq          & $-$1.53 & $-$0.079 & $<$0.001 \\
Saudi Arabia vs.\ Jordan & $+$1.09 & $+$0.056 & $<$0.001 \\
Saudi Arabia vs.\ Syria  & $+$1.36 & $+$0.068 & $<$0.001 \\
Saudi Arabia vs.\ Iraq   & $-$1.30 & $-$0.063 & $<$0.001 \\
Jordan vs.\ Iraq         & $-$2.39 & $-$0.122 & $<$0.001 \\
Syria vs.\ Iraq          & $-$2.66 & $-$0.132 & $<$0.001 \\
\hline
\end{tabular}
\end{table}

\autoref{fig:violins} shows that negativity is not merely the modal tendency
but the dominant mode across all five full distributions. Iraq's negativity
violin shows the widest spread and the longest upper tail. Jordan and Syria are
visually nearly identical in distributional shape---confirmed by their
statistical non-significance (Bonferroni $p = 1.0$). Yemen and Saudi Arabia are
also statistically indistinguishable in negativity, despite Yemen's active armed
conflict.

\subsection{Communal solidarity and singular grief}
\label{sec:rq2}

RQ2 studies the emotional register underlying the negative sentiment. The 28-category emotion analysis reveals an affective landscape that is
simultaneously communal at the regional level and nationally differentiated in
specific emotions. With this, there are three key findings.

First, caring and admiration dominate across all five countries. All
emotion scores reported here are length-normalized (\autoref{eq:emotion_norm}).
The highest mean emotion score is caring in Saudi Arabia ($M = 31.6$) and Yemen
($M = 24.1$). Admiration ranks second in all countries, with Yemen highest
($M = 15.22$) and Saudi Arabia a close second ($M = 15.07$). This dominance of
communal, solidarity-oriented emotions within a predominantly negative sentiment
context holds
identically whether emotion scores are length-normalized or left as raw
trigram sums (\hyperref[supp:complength]{S1 Table}). Thus is the benefit of finer-grained emotion analysis: the communal solidarity register is invisible to sentiment analysis alone, which only reports the overall negativity of the corpus.

Second, grief singularizes Iraq. Iraq's mean grief score ($M = 10.03$)
is 5.4 times that of Yemen ($M = 1.87$). That is the largest cross-country ratio
observed for any single emotion, and larger still than the corresponding
4.2-fold ratio obtained without length normalization. Iraq also shows the
highest anger ($M = 7.19$), followed closely by Syria ($M = 6.12$) and Jordan
($M = 5.85$).

Third, optimism is paradoxically elevated in Yemen ($M = 10.26$),
Syria ($M = 7.58$), and Iraq ($M = 7.58$). All
three countries were experiencing severe ongoing conflict. This may reflect
aspirational commentary (resistance narratives, calls for liberation,
post-conflict reconstruction) coexisting structurally with negative evaluative
sentiment without contradicting it.

\begin{figure}[!ht]
\centering
\includegraphics[width=\linewidth]{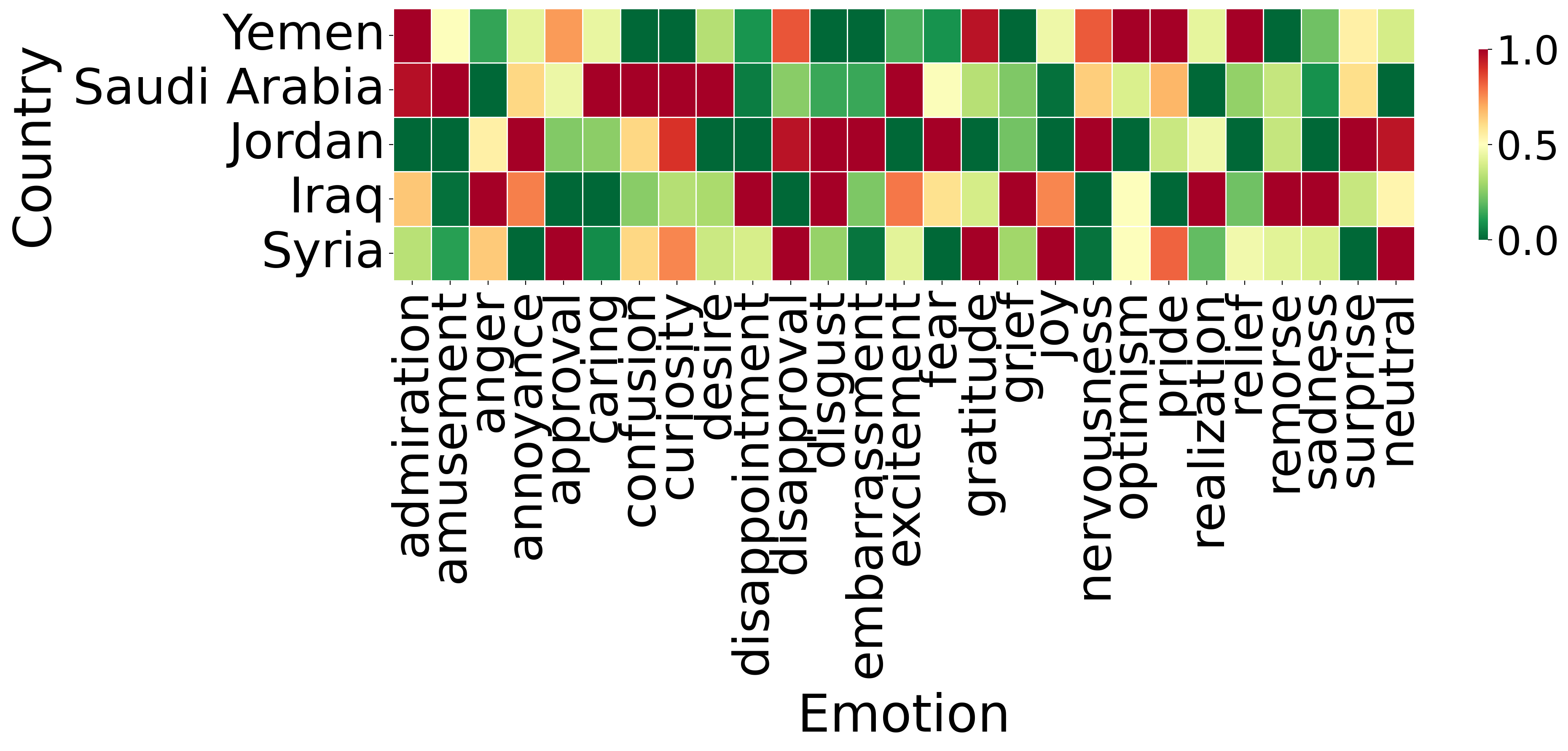}
\caption{{\bf Normalized mean emotion profiles across five countries.}
Cell values are scaled 0--1 within each emotion column to highlight relative
cross-country differences. Caring and admiration dominate throughout. Grief differentiates Iraq from the other four countries. Optimism is
elevated in Yemen and Syria relative to the Gulf and Levant.}
\label{fig:heatmap}
\end{figure}

\begin{figure}[!ht]
\centering
\includegraphics[width=\linewidth]{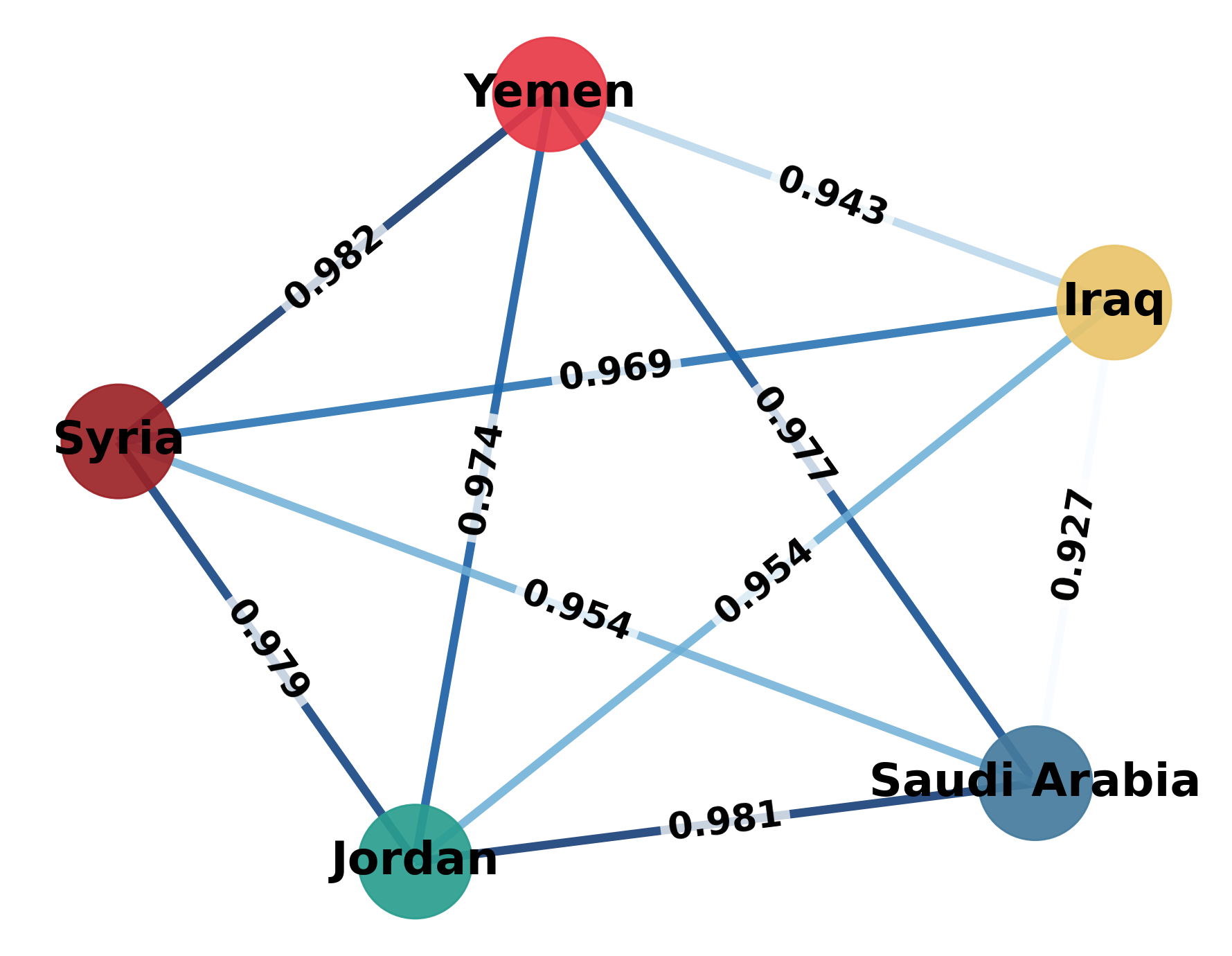}
\caption{{\bf Country emotion-profile similarity network.}
Nodes represent countries; edge weights are cosine similarities between mean
28-dimensional, length-normalized emotion vectors $\mathbf{E}_c$.
Thicker/darker edges indicate greater similarity. Yemen--Syria is the most
similar pair (0.982), closely followed by Saudi Arabia--Jordan (0.981);
Saudi Arabia--Iraq is the least similar (0.927).}
\label{fig:network}
\end{figure}

\begin{figure}[p]
\centering
\begin{minipage}[t]{0.49\linewidth}
  \centering
  \includegraphics[width=\linewidth,height=2.45in,keepaspectratio]{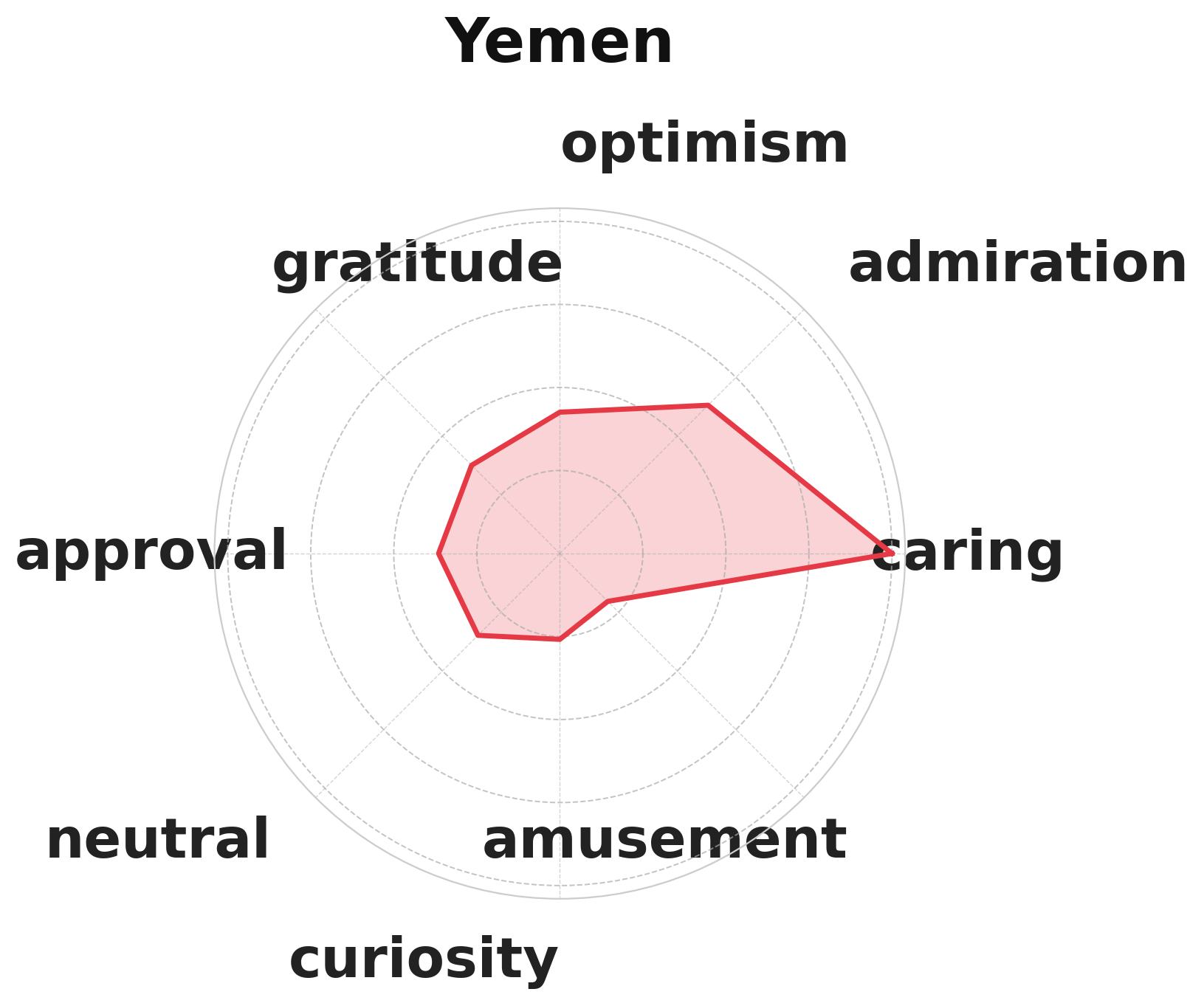}\\
  \small{(A) Yemen}
\end{minipage}
\hfill
\begin{minipage}[t]{0.49\linewidth}
  \centering
  \includegraphics[width=\linewidth,height=2.45in,keepaspectratio]{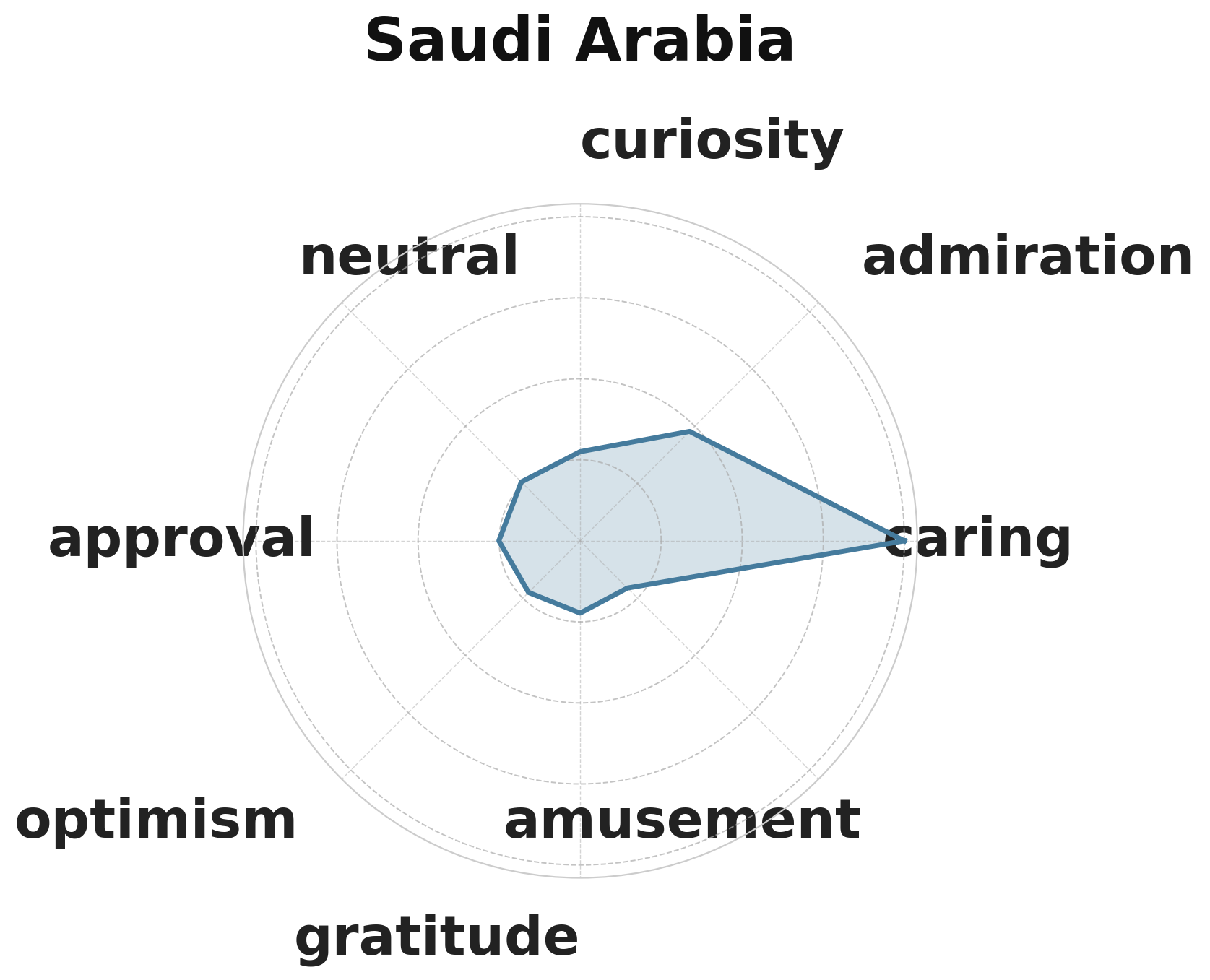}\\
  \small{(B) Saudi Arabia}
\end{minipage}

\vspace{0.2em}

\begin{minipage}[t]{0.49\linewidth}
  \centering
  \includegraphics[width=\linewidth,height=2.45in,keepaspectratio]{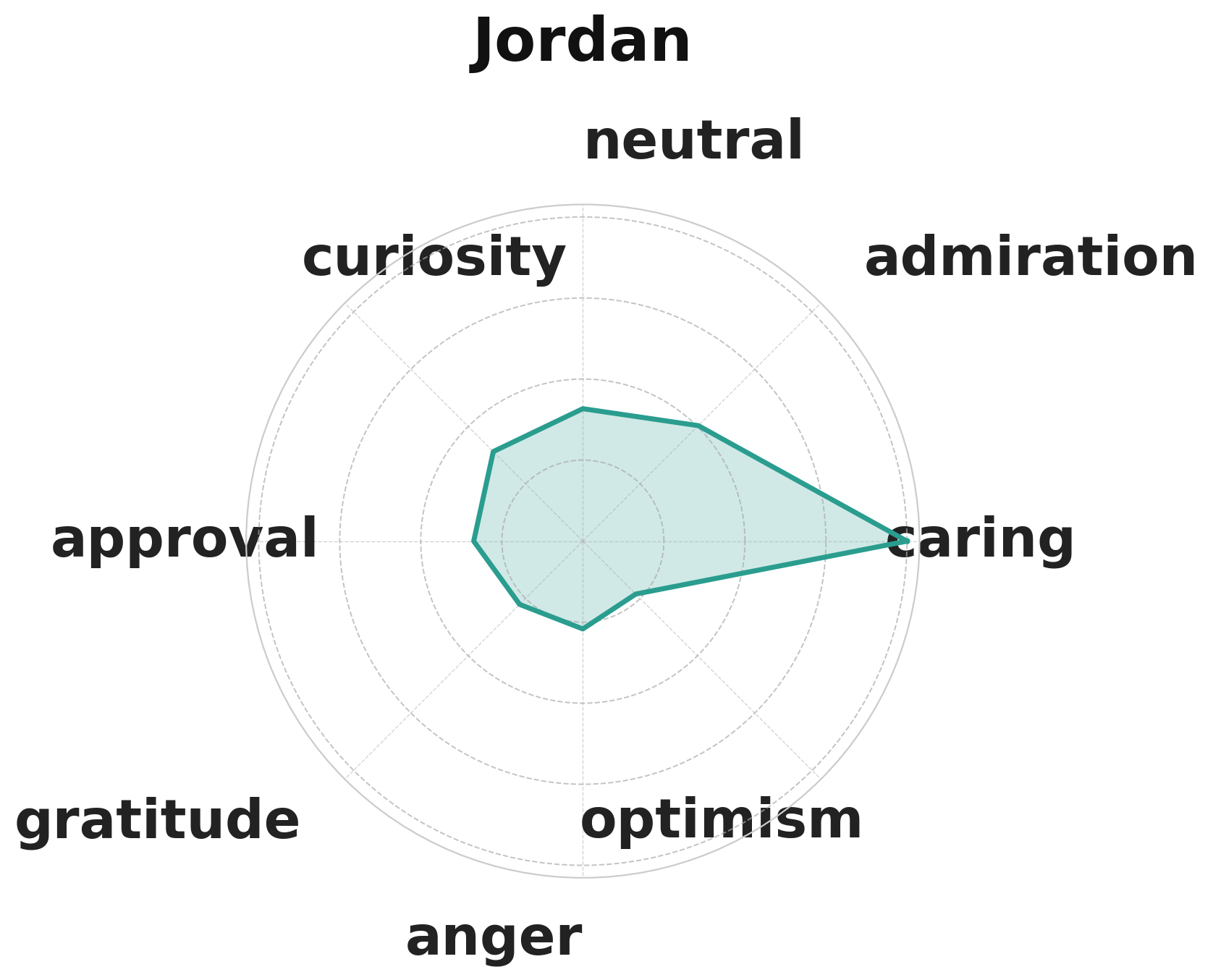}\\
  \small{(C) Jordan}
\end{minipage}
\hfill
\begin{minipage}[t]{0.49\linewidth}
  \centering
  \includegraphics[width=\linewidth,height=2.45in,keepaspectratio]{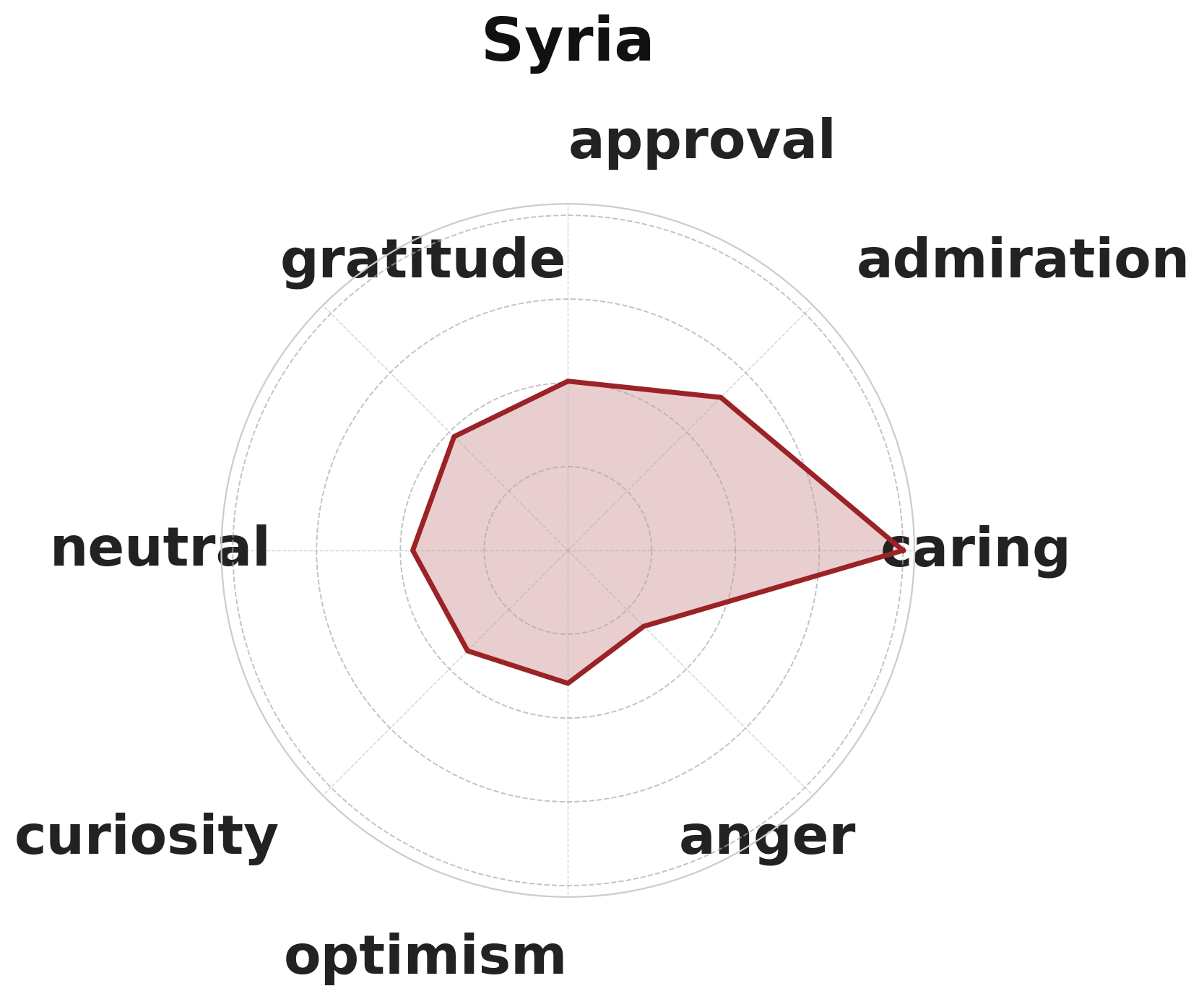}\\
  \small{(D) Syria}
\end{minipage}

\vspace{0.2em}

\begin{minipage}[t]{0.49\linewidth}
  \centering
  \includegraphics[width=\linewidth,height=2.45in,keepaspectratio]{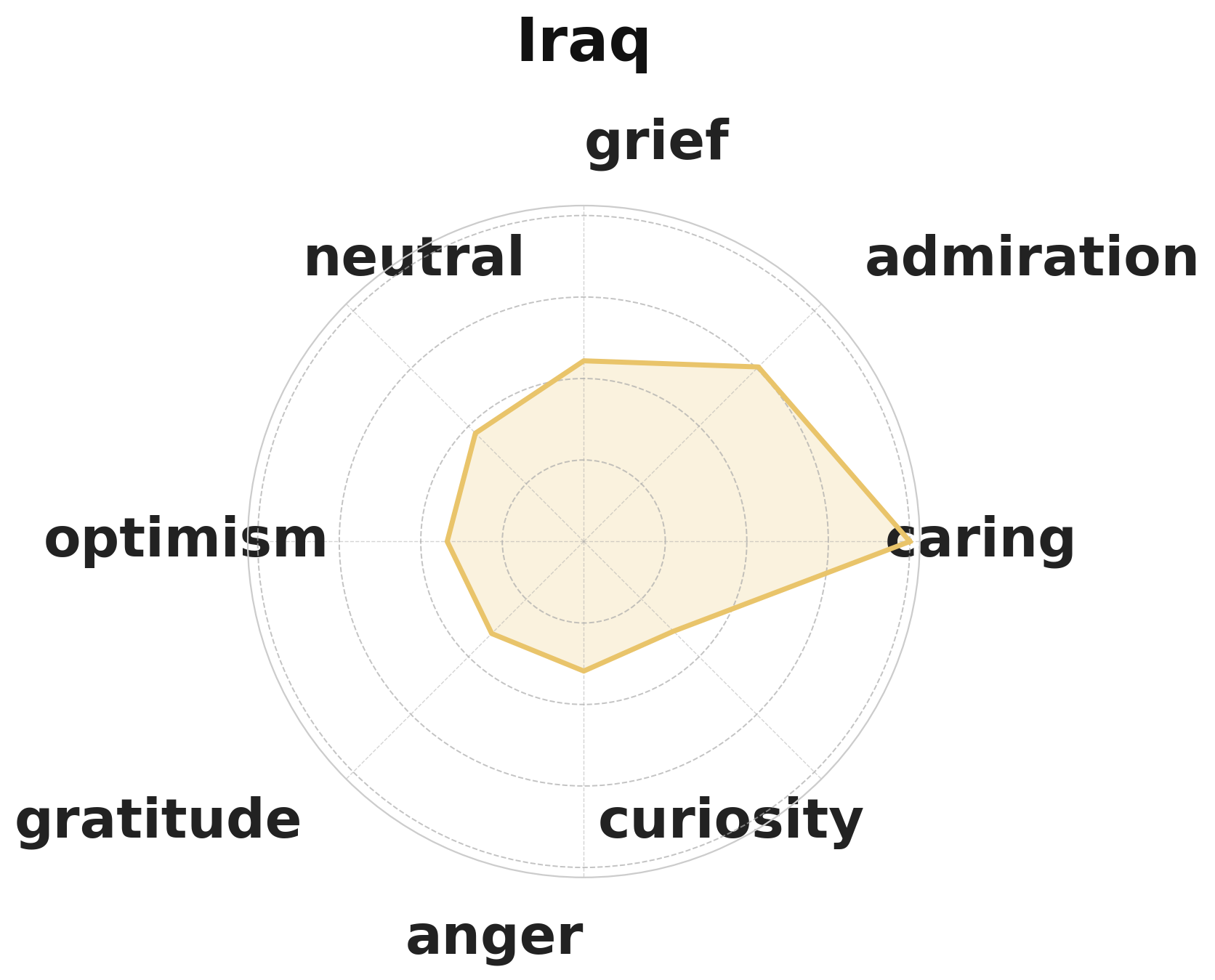}\\
  \small{(E) Iraq}
\end{minipage}

\caption{{\bf Emotion radar charts for each of the five countries
(top-10 emotions, normalized to country maximum).}
(A) Yemen: admiration, optimism, and caring dominate.
(B) Saudi Arabia: caring and admiration peak highest with suppressed anger.
(C) Jordan: high caring alongside elevated anger and curiosity.
(D) Syria: optimism and caring despite ongoing conflict.
(E) Iraq: grief and anger disproportionately elevated while admiration remains
high.}
\label{fig:radar}
\end{figure}

\autoref{fig:network} presents the similarity network of country emotion profiles (full statistics in \autoref{tab:cosine}). All pairwise cosine similarities of country emotion profiles exceed 0.927, confirming a shared affective grammar that
transcends national borders. The most similar pair is Yemen--Syria (0.982),
narrowly ahead of Saudi Arabia--Jordan (0.981). The least similar pair is
Saudi Arabia--Iraq (0.927), consistent with their contrasting political
economies, media environments, and relationships with US military presence.
Country-level radar
charts (\autoref{fig:radar}) illustrate the distinctive emotional fingerprint of
each country. The Yemen-oriented corpus is characterized by admiration, optimism, and caring. Saudi
Arabia shows the fullest polygon of the five countries: caring and admiration
peak highest, with anger lowest, consistent with a state-managed media
environment that may constrain oppositional expression. Jordan combines high
caring with elevated anger and curiosity. Syria shows elevated optimism and
caring despite civil war, potentially reflecting diaspora commentary. Iraq shows
the most distinctive profile: grief and anger are disproportionately elevated
while admiration remains high, consistent with mourning directed at civilian
loss combined with reverence for resistance figures.

\subsection{US-related content produces a directional reversal by country}
\label{sec:rq3}
RQ3 looks at the relationship between US-related content and sentiment.
Our results show that the relationship between US-related content and sentiment is not uniform. Its direction differs across the five country-oriented corpora. In Yemen, Saudi Arabia, and Jordan, USA-mentioned comments are significantly less negative than non-USA comments. In contrast, in Syria and Iraq, USA-mentioned comments are significantly more negative than non-USA comments. All ten within-country comparisons (Positivity and Negativity
for each of five countries) are statistically significant (Mann-Whitney $U$;
all $p < 0.001$). \autoref{tab:usa} presents the full results.

\begin{table}[!ht]
\caption{
{\bf USA-mentioned vs.\ non-USA comments: mean Negativity and Mann-Whitney $U$
test results by country.}
$\Delta$ = USA mean minus non-USA mean (pp = percentage points). All
$p < 0.001$.}
\label{tab:usa}
\begin{adjustwidth}{-2.25in}{0in}
\begin{tabular}{lrrrrrlr}
\hline
\textbf{Country} & \textbf{\textit{n} USA} & \textbf{\textit{n} non-USA} &
\textbf{Neg.\ USA} & \textbf{Neg.\ non-USA} &
\textbf{$\Delta$ (pp)} & \textbf{Direction} & \textbf{\textit{p}} \\
\hline
Yemen        & 8,192 & 10,828 & 81.9\% & 84.1\% & $-$2.2 & USA less negative & $<$0.001 \\
Saudi Arabia & 2,733 & 16,268 & 81.0\% & 83.8\% & $-$2.8 & USA less negative & $<$0.001 \\
Jordan       & 4,312 &  5,961 & 80.5\% & 83.6\% & $-$3.0 & USA less negative & $<$0.001 \\
Syria        & 3,708 &  5,402 & 84.3\% & 80.5\% & $+$3.8 & USA more negative & $<$0.001 \\
Iraq         & 4,795 &  5,526 & 85.7\% & 83.8\% & $+$2.0 & USA more negative & $<$0.001 \\
\hline
\end{tabular}
\end{adjustwidth}
\end{table}

There is substantial variation in USA-mention prevalence. Yemen (43.1\%), Jordan
(42.0\%), Iraq (46.5\%), and Syria (40.7\%) all have roughly comparable
USA-mention rates, while Saudi Arabia's rate is markedly lower (14.4\%). The
directional reversal is therefore not an artifact of comparing large- versus
small-USA-mention countries; for example, Syria and Yemen have similar USA-mention
rates but opposite effect directions.

\begin{figure}[!ht]
\centering
\includegraphics[width=\linewidth]{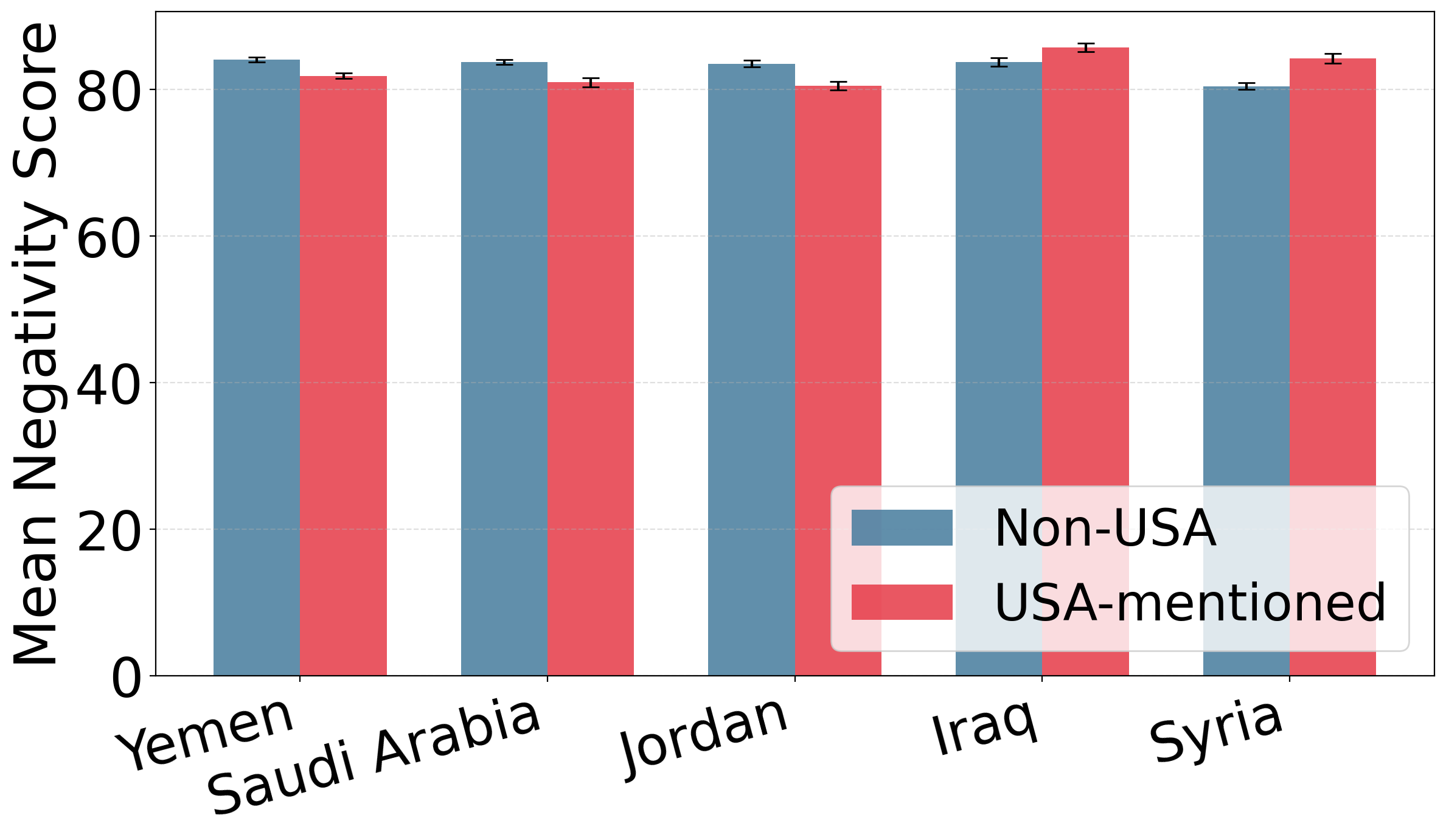}
\caption{{\bf Mean Negativity for USA-mentioned (red) versus non-USA (blue)
comments by country, with 95\% CI error bars.}
The directional reversal is clear: Yemen, Saudi Arabia, and Jordan show lower
negativity in USA-mentioned content; Syria and Iraq show higher negativity. All
differences significant (Mann-Whitney $U$, $p < 0.001$).}
\label{fig:usa}
\end{figure}

% ======================================================================
\section{Discussion}
% ======================================================================

\subsection{Coexistence of negativity and communal solidarity}

Our findings reveal three patterns that, together, are difficult to
account for within a simple grievance model of online political negativity.
A grievance model predicts that high negative sentiment would co-occur
primarily with anger, disapproval, or disgust, since moral-emotional
language of this kind is what drives the spread of negative political
content online \cite{brady2017emotion}. Instead, we observe that caring and
admiration are the dominant emotion categories across all five countries,
coexisting with mean negativity rates of 82--85\%. This co-occurrence is
consistent across independent national corpora organized by different keyword
sets, which makes it unlikely to be an artifact of data collection. The pattern
suggests that Arabic YouTube users express negative political evaluations
alongside communal solidarity orientations simultaneously, rather than one
displacing the other.

A simple anti-Americanism account---one that treats hostility to the US as a
fixed reaction to what the United States \emph{is} rather than what it
\emph{does} \cite{katzenstein2007antiamericanisms}---would predict that
US-related content is more negative than average across all countries.
Instead, we find a directional
reversal: US-mentioned comments are significantly less negative in Yemen, Saudi
Arabia, and Jordan, but significantly more negative in Syria and Iraq. In this
five-country sample, the difference between these two clusters does not map onto linguistic, sectarian,
or ideological divisions in an obvious way, but does correspond to the presence
or absence of active US military operations.

Cross-national emotion profile similarity (cosine similarity 0.927--0.982)
across five politically distinct countries suggests that the emotional structure
of Arabic YouTube political discourse is organized at a regional level, not only
a national one. This level of convergence across independent corpora is
consistent with the existence of a shared affective grammar. That is, recurring
emotional orientations that appear to transcend national borders even in the
absence of cross-national organizational coordination, echoing earlier
accounts of a pan-Arab public sphere sustained by shared media exposure
rather than by any single national institution \cite{lynch2006voices}.

\subsection{Affective publics as an interpretive framework}
\label{sec:affpub}

The concept of \emph{affective publics}, as developed by Papacharissi
\cite{papacharissi2015}, provides the interpretive frame for the
co-occurring patterns described in the previous section. Papacharissi's account holds that digital platforms
increasingly constitute political publics through shared emotional registers
rather than through deliberative exchange. An ambient public is assembled less by
shared argument than by a shared structure of feeling, capable of holding
users together around a political moment without formal coordination or
ideological consensus. The five-country corpus is broadly consistent with this
account. Our results show thathigh negative sentiment coexists with communal solidarity emotions,
and that coexistence holds independently across five nationally distinct
corpora rather than emerging from a single shared discourse community.

With this frame, negativity is not a departure from the affective
public but one of its constitutive registers. Sentiment expressed online
functions as political texture rather than noise---a signal of how users are
affectively positioned toward events, actors, and institutions, carried
through evaluative valence as much as through argument. The overwhelming
negativity documented above (\autoref{sec:rq1}) reflects this evaluative layer of
the affective public: users registering political dissatisfaction. The
accompanying dominance of caring and admiration reflects its solidary layer:
orientation toward co-participants and shared political causes, expressed
alongside the negativity rather than displacing it. The same comment can carry
both registers because they answer different questions. For example, negativity answers
how a user evaluates a given actor or event; caring and admiration answer who
the user is affectively aligned with.

Within this framework, affective intensity does not distribute evenly across
political content; it concentrates around specific causes that come to
organize a public's attention over time. The corpus is consistent with this:
negativity and solidarity emotions alike cluster around a small set of
recurring political causes in each country (i.e., Gaza solidarity, the
Houthi-Saleh civil war inheritance in Yemen, the US-led invasion and its
aftermath in Iraq) rather than distributing uniformly across every political
topic an affective public might in principle attend to. Political causes, in
other words, function as the organizing structure around which the affective
public accumulates its emotional charge, rather than affect attaching evenly
to political discourse in general.

If affective publics organize around shared causes rather than shared
nationality per se, and several of the causes identified above (i.e., the Gaza
war, US regional presence) are common across the five corpora, the framework
predicts convergent emotional registers wherever these shared causes dominate
local discourse, independent of any cross-national coordination. This is what
the data show: the five country-level emotion profiles are structurally
similar (cosine similarity 0.927--0.982; \autoref{sec:rq2}), with caring and
admiration dominant in every corpus and the same secondary structure of
anger, curiosity, and grief recurring throughout. Under the affective publics
account, this convergence indicates a shared regional structure of feeling
organized around overlapping political causes, not a single undifferentiated
Arabic-language public; the country-specific departures examined below
(\autoref{sec:crossnational}) remain fully compatible with this shared underlying register.

The US-sentiment reversal (\autoref{sec:rq3}) offers a further test of the same
logic. If an affective public's negativity is organized around specific,
lived political causes rather than fixed identity categories, the same
referent---the United States---should not carry a uniform valence across
corpora unless it names the same lived cause everywhere. It does not:
US-related comments are less negative than average in Yemen, Saudi Arabia,
and Jordan, where US engagement is indirect, but more negative in Syria and
Iraq, where it more direct. Under the affective publics framework, this is
consistent with users in each corpus evaluating ``the United States'' not as
an abstract political category but as it is locally instantiated---as an
object bound up with the specific causes that organize each corpus's
affective register (\autoref{sec:namedactors} examines this further through the named
entities each corpus attaches to those causes).

\subsection{Cross-national emotional convergence and regional affective structure}
\label{sec:crossnational}

The pairwise cosine similarities between country-level emotion profiles
(0.927--0.982) are notable given that the five corpora were collected
independently using country-specific keyword sets organized around distinct
national political contexts. This degree of convergence suggests that the
emotional structure of Arabic YouTube political discourse reflects regional
organizing principles rather than purely national ones.

Within the affective publics framework introduced above (\autoref{sec:affpub}), this
pattern is consistent with a single, loosely bounded regional public rather
than five separate national ones. An affective public
does not require formal organization or national containment. Instead, an affective public holds
together through shared exposure to overlapping political causes and a common
media ecology, not through membership or coordination. The Gaza conflict,
regional instability, and the recurring presence of the United States as a
political actor are causes that plausibly reach audiences across all five
corpora simultaneously, which would be sufficient under this account to
produce convergent emotional registers without any cross-national
organizational structure linking the five discourse communities.

It is important to note what this result does and does not indicate. A cosine
similarity of 0.927 between Saudi Arabia and Iraq reflects structural similarity
in emotional profile shape, not identity. The shared template, in which caring
and admiration dominate while anger and grief remain secondary, accommodates
country-specific deviations. Of note are Iraq's elevated grief and Jordan's elevated
anger, without these differences eliminating the overall convergence. This finding
is one of regional emotional structure, not emotional uniformity across contexts.

\subsection{Named actors as affective anchors}
\label{sec:namedactors}

The named entity data (specific details in \autoref{supp:entities}) add a dimension to the affective
grammar finding that aggregate sentiment and emotion scores cannot capture: the
specific political actors around whom affect is organized, and whether those
actors are current or historical. The pattern that emerges is not one of pure
present-tense political reaction but of an affective public in which historical
memory and contemporary events are layered, with some historical figures
circulating across national borders as shared reference points.

In Yemen, the prominent actors are Abdul-Malik al-Houthi (the current Houthi
leader) and the former President Ali Abdullah Saleh (killed by Houthi forces
in 2017) and his nephew Tariq Saleh. The presence of both Houthi and Saleh
figures in the same corpus reflects the unresolved political inheritance of the
civil war \cite{lackner2017yemen}. Notably, Mohammed bin Salman also appears prominently in Yemeni
discourse, reflecting Yemen's inescapable entanglement with Saudi intervention.
The pan-Arab nationalist resonance of Abdel Nasser invoked in the comments with caring and
admiration, the discourse's dominant emotions suggests that contemporary
solidarity is inflected with a longer historical longing for a unified Arab
political project. 

Iraq's named actor profile anchors the grief singularity in specific historical
trauma. Saddam Hussein, Tariq Aziz (his foreign minister), and Ali Hassan
al-Majid (or ``Chemical Ali''), architect of the Anfal campaign are the most
prominent figures. These are not actors producing current political affect. Rather, they
are figures associated with mass atrocity and the subsequent US-led invasion that
dismantled the Ba'athist state \cite{tripp2007history}. That they dominate Iraqi YouTube discourse
twenty years later suggests that Iraqi grief is not simply about present violence
but about an accumulated historical wound in which the US was a central actor,
which contextualizes why Iraq's US-related content is the most negative of the
five countries. Al-Sisi's (the Egyptian president
with no direct Iraq connection) appearance in Iraqi discourse  may reflect a broader regional strongman frame
through which Iraqis process political authority.

In Jordan, the most prominent named figures are historical: Wasfi al-Tal
(Prime Minister assassinated by the PLO in 1971), King Hussein (deceased), and
Yasser Arafat \cite{robins2004history}. Their appearance alongside King Abdullah and current political
discourse signals that Jordanian political affect is organized substantially
through the Palestinian cause as a multigenerational historical frame, consistent
with the country's geographic proximity to Gaza and the demographic composition
of its population. The shared presence of Yasser Arafat in both Jordanian and
Syrian corpora connects the two countries' discourse to a common Palestinian
referent.

Saudi discourse is distinctive in naming Vladimir Putin. Putin is the only non-MENA
international figure to appear prominently in any country corpus. In summer 2024,
during simultaneous Russia-Ukraine and Israel-Gaza conflicts, Putin's appearance
likely reflects a comparative frame (``why does the West sanction Russia for
what it does in Ukraine but support Israel in Gaza?'') that circulated widely in
MENA discourse. Saddam Hussein's (the figure
Saudi Arabia fought against in the Gulf War) appearance in Saudi discourse suggests that historical figures
function as negative reference points in Saudi political memory, alongside the
more predictable domestic figures of MBS and his predecessors.

These named entity data suggest that the shared emotional
structure documented in the emotion profiles may not be organized only around
present political events. Historical trauma figures (i.e., Saddam Hussein in Iraq,
Wasfi al-Tal and Yasser Arafat in Jordan) appear as prominent referents
alongside current actors, suggesting that historical memory is an active
component of political discourse in Arabic YouTube comment sections, not merely
historical background. Within the affective publics framework, this is
consistent with Papacharissi's \cite{papacharissi2015} account of networked
storytelling, in which publics narrate and sustain themselves by threading
present events into a continuous, affectively charged story rather than
treating each political moment as a discrete, self-contained occurrence;
historical figures function here as recurring anchors in that ongoing
narrative rather than as settled background.

\subsection{Geopolitical context and the US-sentiment reversal}

The directional reversal in the US-sentiment relationship---less negative in
Yemen, Saudi Arabia, and Jordan; more negative in Syria and Iraq---does not
correspond to obvious differences in ideology, sect, or political regime type
across the five countries. Among the contextual differences considered here, the most apparent distinction between
the two clusters is the presence of active US military operations. US forces
maintain a direct operational presence in northeastern Syria and at Iraqi
military installations, while US engagement in Yemen and the Gulf operates
primarily through indirect support and alliance frameworks.

One interpretation consistent with these findings, and with the affective
publics account developed above (\autoref{sec:affpub}), is that the valence of
US-related content is conditioned not by abstract orientations toward
``America'' as a category. Instead, it is conditioned by the specific, locally experienced political
cause that ``the United States'' names in each corpus. If this account is
correct, comments referencing the US in Syria and Iraq may activate the same
negative emotional register as domestic conflict content, because US actions
are experienced as a proximate part of the same ongoing cause. In countries
where US engagement is materially indirect, the same referent may not
activate the same emotional intensity.

This interpretation is consistent with the broader pattern documented in
\autoref{sec:affpub}, in which the affective public's emotional register tracks lived
political causes rather than fixed ideological categories. However, the
cross-sectional design of this study cannot rule out alternative explanations,
including differences in the demographic composition of YouTube users across the
five countries, differences in the specific videos generating US-related
comments, or topic confounding in the keyword-based collection approach. While the
directional reversal is a robust empirical finding; its causal interpretation
remains tentative and should be examined in future work with more controlled
designs.

\subsection{Research and Policy implications}

% These results carry two kinds of implications: one for how digital political
% communication should be studied, and one for how it should be acted on.

Methodologically, sentiment and emotion prove to be complementary analysis. Sentiment scores alone place Arabic YouTube political discourse
at the negative end of the spectrum. The emotion data, however, reveal that this negativity coexists with a communal one, where caring and admiration dominate the emotional register, suggesting that studies of affectiev publics can consider both sentiment and emotion to capture the full affective orientation of political discourse. More broadly, the findings support treating the affective publics framework
\cite{papacharissi2015} as operational rather than purely interpretive: its
core constructs (ambient, non-organizational publics; affectively charged
causes as organizing objects, and networked storytelling across present and
historical events) can be analyzed computationally at scale, although this
study's cross-sectional design cannot confirm the causal mechanisms the
framework proposes.

The policy implications follow the same logic. Since the emotional register surrounding US-related content differs across the five country-oriented corpora, communication strategies premised on a single regional response to US foreign policy may not fit the audience they are meant for. Counter-messaging pegged to aggregate MENA negativity figures can misfire: the US-sentiment reversal shows that the same referent carries opposite valence depending on local political context, so a message calibrated to the region's average register will be miscalibrated for every country in particular. In Syria and Iraq, where US-related content is more negative and the dominant register is grief-and-anger, messaging that foregrounds shared humanitarian concern and recognition of civilian suffering is more likely to land than policy-centered argument; in Yemen, Saudi Arabia, and Jordan, where the register runs more favorable, that same humanitarian framing would be answering a question no one is asking. There should therefore be country-level strategies tailored to each case's emotional register and geopolitical context, not a one-size-fits-all regional message.

Since the emotional register shifts systematically with geopolitical events rather
than holding fixed, it is also a usable signal rather than only a
retrospective one. The pipeline described here is technically capable of
running in near real time, which suggests a further application:
integrating sentiment-and-emotion monitoring into open-source intelligence
workflows as an early-warning indicator of affective shifts tied to
unfolding events, rather than solely performing retrospective analysis.

\subsection{Limitations and Future Work}
Several limitations bound the scope of these findings. 
The use of CAMeL-Lab BERT and EmoRoBERTa for sentiment and emotion classification, respectively, introduces potential biases. CAMeL-Lab BERT is trained specifically on dialectal Arabic generally but not equally across all regional varieties. While the model is state-of-the-art for Arabic sentiment classification, future work should incorporate explicit dialect identification to account for potential systematic differences in model performance across Yemeni, Gulf, Levantine, and Iraqi dialects. EmoRoBERTa requires English input via machine translation, which may suppress culturally specific vocabulary for religious sentiment, collective mourning, and political solidarity. Therefore, emotion results should be read as relative cross-country comparisons rather than absolute prevalence estimates, a limitation compounded by the absence of any Arabic-native benchmark for the full 28-category GoEmotions taxonomy. Finally, the use of YouTube discourse self-selects vocal participants rather than representative national populations, and country assignment derives from search context rather than verified user location or nationality. Future work should examine whether the patterns observed here replicate on other Arabic-language platforms including Instagram, TikTok, and Telegram, and whether emotion profiles differ by comment-level features such as thread depth, video genre, or channel type.

% ======================================================================
\section{Conclusion}
% ======================================================================

This study analyzed 67,725 Arabic YouTube comments collected around
socio-political topics associated with Yemen, Saudi Arabia,
Iraq, Jordan, and Syria to characterize the sentiment and emotion structure of
political discourse on the platform across five MENA countries. The findings of our cross-country analysis, when interpreted through the affective publics framework provides a useful resource for understanding Arabic-language MENA digitial discourse. Our analysis yield three key findings. 

First, Sentiment is predominantly negative across all five countries, with Iraq showing the
highest negativity and Syria and Jordan statistically indistinguishable from one
another. Despite this pervasive negative sentiment, the most frequently observed
emotion categories are caring and admiration, which are communal solidarity orientations
rather than anger or disgust. This pattern is consistent across all five
independently collected national corpora, suggesting that negative political
evaluation and communal affective solidarity co-occur in Arabic YouTube discourse.

Second, country-level emotion profiles are structurally similar across all five countries, consistent with the existence of a
shared regional affective structure in Arabic YouTube political discourse. Iraq
departs from the regional template most substantially in grief, which is consistent with Iraq's distinct
accumulation of mass-violence events and the anchoring of conversations to historical figures associated with atrocity.

Third, US-related comments are significantly less negative in Yemen, Saudi Arabia, and
Jordan but significantly more
negative in Syria and Iraq. This
directional reversal corresponds with the distinction between countries where US
military presence is direct and countries where it is indirect, and suggests
that the relationship between foreign power references and sentiment negativity
in MENA digital discourse is geopolitically conditioned rather than uniform.

% ======================================================================
\section*{Declarations}
% ======================================================================

\paragraph{Author contributions.}
L.H.X.N.: Methodology, Analysis,Writing---Review \& Editing.
C.D.A.: Conceptualization, Funding Acquisition, Project Administration,
Writing---Original Draft, Writing---Review \& Editing.
A.M.: Data Curation, Methodology, Software, Formal Analysis.
A.A.: Methodology, Software, Validation, Writing---Review \& Editing.
L.Y.H.: Conceptualization, Writing---Review \& Editing, Supervision.

\paragraph{Competing interests.}
The authors declare no competing interests. This research was funded by the
U.S.\ Office of Naval Research (award N000142212549). 

\paragraph{Data availability.}
The anonymized comment dataset and the video data can be made available upon request, with accordance to YouTube's Terms of Service.

% ======================================================================
\section*{Acknowledgments}
% ======================================================================

The authors wish to thank Dr.\ Gregg R.\ Murray for his review, editing, and
thoughtful suggestions on earlier versions of this manuscript. The authors also thank Chheten Sherpa for his research assistance in this project.

\bibliography{PLOSTwo_V2}

\clearpage
\appendix

\renewcommand{\thefigure}{S\arabic{figure}}
\renewcommand{\thetable}{S\arabic{table}}
\renewcommand{\thesection}{S\arabic{section}}

\setcounter{figure}{0}
\setcounter{table}{0}
\setcounter{section}{0}

\section{Search Keywords}
\label{supp:keywords}

The full list of Arabic search terms and their English translations, organized
by country and thematic domain, used for YouTube data collection during summer
2024. Searches were conducted using the \texttt{youtubesearchpython} library.
Keywords were developed by native-speaker members of the research team with
expertise in MENA regional politics, and validated against pilot results before
final data collection.

\section{Sentiment Scores by Country}
\label{supp:sentiment}

\begin{table}[!ht]
\caption{
{\bf Mean (Standard Deviation) sentiment scores by country (0--100 scale).}
ANOVA: Negativity $F(4, 67{,}720) = 29.36$, $p < 0.001$;
Positivity $F(4, 67{,}720) = 40.52$, $p < 0.001$.
Bold indicates highest mean negativity.}
\label{tab:sentiment}
\begin{adjustwidth}{-1.5in}{0in}
\begin{tabular}{lrrrr}
\hline
\textbf{Country} & \textbf{\textit{n}} & \textbf{Mean Positivity (SD)} &
\textbf{Mean Negativity (SD)} & \textbf{Mean Neutrality (SD)} \\
\hline
Yemen        & 19,020 & 21.46 (6.87)  & 83.16 (18.19)          & 0.893 (0.045) \\
Saudi Arabia & 19,001 & 22.29 (8.82)  & 83.39 (20.47)          & 0.892 (0.051) \\
Jordan       & 10,273 & 22.14 (7.37)  & 82.29 (18.45)          & 0.894 (0.044) \\
Syria        &  9,110 & 22.31 (7.93)  & 82.03 (19.58)          & 0.895 (0.035) \\
Iraq         & 10,321 & 22.43 (7.83)  & \textbf{84.69 (20.64)} & 0.890 (0.052) \\
\hline
\end{tabular}
\end{adjustwidth}
\end{table}

\begin{table}[!ht]
\caption{
{\bf Pairwise cosine similarities between mean 28-dimensional, length-normalized
emotion profile vectors ($\mathbf{E}_c$).}
All values exceed 0.927, indicating a shared regional affective grammar.
Yemen--Syria is the most similar pair; Saudi Arabia--Iraq is the least similar.}
\label{tab:cosine}
\begin{tabular}{lrrrrr}
\hline
& \textbf{Yemen} & \textbf{Saudi Arabia} & \textbf{Jordan} &
  \textbf{Syria} & \textbf{Iraq} \\
\hline
\textbf{Yemen}        & ---   & 0.977 & 0.974 & 0.982 & 0.943 \\
\textbf{Saudi Arabia} & 0.977 & ---   & 0.981 & 0.954 & 0.927 \\
\textbf{Jordan}       & 0.974 & 0.981 & ---   & 0.979 & 0.954 \\
\textbf{Syria}        & 0.982 & 0.954 & 0.979 & ---   & 0.969 \\
\textbf{Iraq}         & 0.943 & 0.927 & 0.954 & 0.969 & ---   \\
\hline
\end{tabular}
\end{table}

\section{Named Persons by Country}
\label{supp:entities}

\begin{figure}[!ht]
\centering
\includegraphics[width=\linewidth]{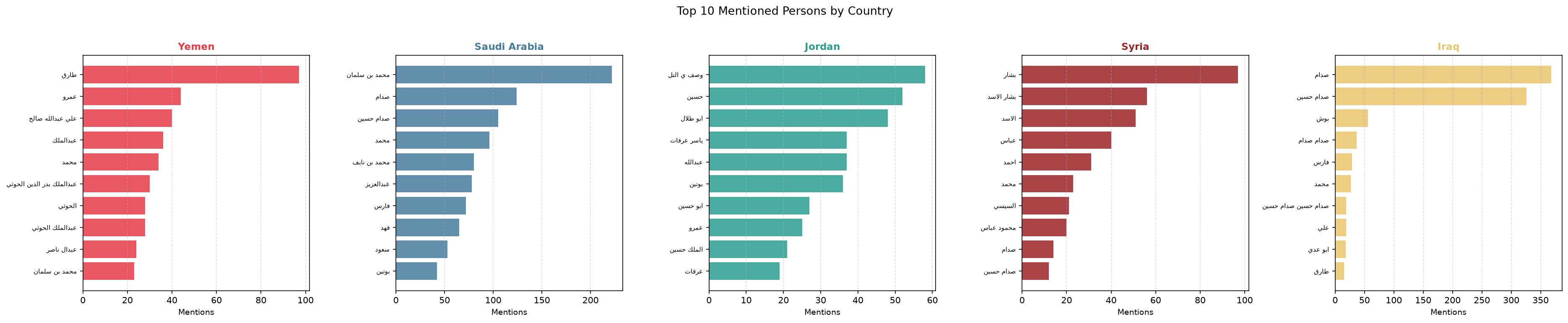}
\caption{{\bf Most frequently mentioned named persons by country,} extracted
from NER-annotated comment text. Political leaders and resistance figures
dominate; cross-country overlap in mentioned actors reflects the regionally
integrated nature of Arabic YouTube political discourse.}
\label{fig:entities}
\end{figure}

\section{Comment Length by Country}
\label{supp:complength}

Comment-level emotion scores (\autoref{eq:emotion_norm}) are sums over
overlapping trigrams and therefore scale with comment length. \autoref{tab:complength}
reports the token-count distribution by country underlying the length
normalization described in the Computational pipeline subsection. Country
means differ at conventional significance ($F(4, 67{,}720) = 9.66$, $p <
0.001$), but the substantive magnitude of this difference is small: the median
comment is 10 tokens in every country, and means range from 13.2 (Syria) to
14.2 (Jordan) tokens---a spread of one token against a standard deviation of
roughly 12. 3{,}411 comments (5.0\%) have fewer than 3 tokens, cannot form a
trigram, and are excluded from the length-normalized emotion analysis
throughout the paper.

As an ablation test, also compared the non length-normalized emotion profiles to the length-normalized ones reported in the main text.
The three headline patterns are unchanged: caring and
admiration remain the top two emotions in all five countries under both
scorings; Iraq's grief elevation relative to Yemen is, if anything, larger after
normalization (raw ratio 4.2$\times$, normalized ratio 5.4$\times$); and
cross-country cosine similarities remain uniformly high (raw range
0.940--0.984; normalized range 0.927--0.982). The one substantive change is
which pair is most similar: under length normalization, Yemen--Syria (0.982)
narrowly overtakes Syria--Iraq (0.969), which was most similar under the raw
scoring; Saudi Arabia--Iraq remains the least similar pair under both scorings.

\begin{table}[!ht]
\caption{{\bf Comment length (token count) by country.}}
\label{tab:complength}
\begin{tabular}{lrrrrrr}
\hline
\textbf{Country} & \textbf{Mean} & \textbf{Median} & \textbf{SD} & \textbf{Min} & \textbf{Max} & \textbf{N} \\
\hline
Iraq          & 13.3 & 10 & 12.0 & 1 & 102 & 10{,}321 \\
Jordan        & 14.2 & 10 & 12.6 & 1 & 97  & 10{,}273 \\
Saudi Arabia  & 13.5 & 10 & 12.4 & 0 & 97  & 19{,}001 \\
Syria         & 13.2 & 10 & 12.1 & 1 & 95  & 9{,}110 \\
Yemen         & 13.7 & 10 & 12.1 & 0 & 98  & 19{,}020 \\
\hline
\end{tabular}
\end{table}

\nolinenumbers

\end{document}